\documentclass[aps,prl,twocolumn,fleqn,floatfix,nofootinbib]{revtex4}

\usepackage{epsfig}
\usepackage{amsmath,amssymb,amsfonts}

\begin{document}

\title{Longitudinal Broadening of Quenched Jets in Turbulent Color Fields}

\author{A. Majumder}
\affiliation{Department of Physics, Duke University, Durham, NC 27708, USA}
\author{B. M\"uller}
\affiliation{Department of Physics, Duke University, Durham, NC 27708, USA}
\author{S. A. Bass}
\affiliation{Department of Physics, Duke University, Durham, NC 27708, USA}

\date{\today}

\begin{abstract}
The near-side distribution of particles at intermediate transverse momentum, 
associated with a high momentum trigger hadron produced in a high 
energy heavy-ion collision, is broadened in rapidity compared with the jet cone. 
This broadened distribution is thought to contain the energy lost by the progenitor parton
of the trigger hadron. We show that
the broadening can be explained as the final-state deflection 
of the gluons radiated from the hard parton inside the medium by soft, transversely
oriented, turbulent color fields that arise in the presence of plasma instabilities. 
The magnitude of the effect is found to grow with medium size and density and 
diminish with increasing energy of the associated hadron. 
\end{abstract}

\maketitle

The emission of hadrons with large transverse momentum is observed to
be strongly suppressed in central collisions of heavy nuclei at the
Relativistic Heavy Ion Collider (RHIC)~\cite{Adcox:2001jp,Adler:2002xw}. 
The suppression is understood to be caused by final-state rescattering 
of the leading parton in the dense medium produced in such collisions,
causing it to lose energy.  Within the framework of perturbative QCD, the 
leading process of energy loss of a fast parton is gluon radiation induced by 
elastic collisions with color charges in the quasi-thermal 
medium~\cite{Gyulassy:1993hr,Baier:1996kr,Zakharov:1997uu}.
Elastic collisions not followed by radiation may also contribute to the energy
loss~\cite{Mustafa:2004dr}.

In this manuscript, we study the fate of the energy lost by the leading parton of 
the trigger jet within the medium. Measurements by the STAR collaboration 
have shown that the leading hadron is accompanied by a cone-shaped pattern 
of secondary hadrons, which closely resembles the jet cone 
around a hard scattered parton in nucleon-nucleon 
collisions~\cite{Adams:2006tj,Adams:2005ph}. We will refer to this as the 
contribution from vacuum fragmentation  (or ``jet'' yield). 
In addition, the trigger parton is observed to be accompanied on the near 
side by a wide ridge~\cite{Putschke:2007mi} (or pedestal~\cite{Chiu:2005ad}) of secondary hadrons, 
which extends along the beam direction over more than one unit of rapidity 
in both directions. We refer to this as the contribution from in-medium 
fragmentation (or ``ridge'' yield).  The energy flow in the ridge-shaped structure 
grows with participant number and is concentrated in hadrons with transverse 
momenta $p_T < 2$ GeV/c.

The inclusive yield of associated hadrons receives contributions from the 
fragmentation of the leading parton as well as from the radiated gluons~\cite{Majumder:2004pt}. 
The yield of associated hadrons produced in the fragmentation of the hard 
parton after its exit from the medium will be assumed to form the jet yield~\cite{Majumder:2004wh}. 
This naturally explains why the azimuthal and rapidity distribution of this contribution 
is identical to that from a jet of the same final energy produced in a $p+p$ collision. 
Hadrons produced in the fragmentation of the gluons radiated within the 
medium or after absorption of those gluons will be assumed to form the ridge 
yield. In either case, the ridge shape will reflect the kinematic distribution of the
radiated gluons at the end of their passage through the medium.
 
It has been argued that the properties of the ridge must be associated with 
the longitudinal expansion of the medium~\cite{Armesto:2004vz,Chiu:2005ad}.
However, the precise mechanism by which the ridge acquires its peculiar 
shape has not been quantitatively explained. In a recent article, 
Romatschke~\cite{Romatschke:2006bb} has shown that the momentum of a 
transversely propagating heavy quark is preferentially broadened in the 
longitudinal direction by elastic collisions in an expanding medium, 
if the expansion leads to a large anisotropy of the momentum distribution of the 
partons composing the medium. He suggested that this phenomenon could
explain the ridge.  We note, however, that elastic scattering is probably not 
the primary source of energy loss for the light partons which initiate the 
observed ridge. We also note that the momentum anisotropy parameter 
$\xi$ of the expanding medium is related to its shear viscosity $\eta$
by the relation~\cite{Asakawa:2006tc}
$\xi \approx 10\,\eta/(s\tau T)$, 
where $s$ denotes the entropy density, $T$ the temperature, and $\tau$ the
time after the initial hard collision. For times relevant to the propagation of the 
jet through the medium ($\tau \geq 1$ fm/c), the large values of $\xi \sim 10$ 
required to explain the extent of the ridge are difficult to reconcile with the low 
shear viscosity of the medium ($\eta/s \leq 0.3$) required by the data
\cite{Teaney:2003kp,Romatschke:2006bb}.

Here, we propose a different mechanism for the origin of the observed ridge,
also related to the longitudinal expansion of the medium. 
It builds on the recent insight that extended color fields are dynamically 
formed in the expanding medium due to the presence of plasma instabilities. 
Such instabilities have been shown to exist in any quark-gluon plasma 
with an anisotropic momentum 
distribution~\cite{Randrup:2003cw,Romatschke:2003ms,Arnold:2003rq}. 
The nearly boost invariant longitudinal expansion of the matter formed in 
relativistic heavy ion collisions necessarily induces an oblate momentum 
distribution of partons with respect to the beam axis. The plasma instabilities 
lead to the exponential growth of soft modes of the glue field, which ultimately 
saturate due to the nonlinear self-interactions of the Yang-Mills field, resulting 
in a turbulent state of the quasi-thermal quark-gluon 
plasma~\cite{Arnold:2005ef,Arnold:2005qs}. Such turbulent color fields
currently present a successful 
mechanism for the fast equilibration of 
the medium~\cite{Randrup:2003cw,Arnold:2003rq} and serve as a source of an 
anomalous shear viscosity~\cite{Asakawa:2006tc}. This naturally explains
the ideal fluid behavior of the medium. Here we explore the
effect of turbulent color fields on the gluons radiated by a hard parton in the medium.

When the hard parton scatters transverse (in the medium co-moving 
frame) to the beam direction, its accompanying halo of soft radiated gluons 
are deflected by the turbulent color fields as the parton traverses the medium. 
The deflection pattern is not isotropic around the jet axis, because the 
direction of polarization of the instability driven color fields is not totally 
random. Detailed studies of the instability pattern~\cite{Romatschke:2003ms} 
have shown that color magnetic field modes polarized transversely to the 
beam direction exhibit the largest growth rates and, therefore, dominate 
the field configurations at early times. The dominance of transversely 
polarized color-magnetic fields results in the preferential deflection
of transversely propagating partons in the direction of the beam axis,
with a deflection angle that changes inversely with the parton momentum.
The resulting pattern will be one in which gluons of moderate $p_T$ 
from the jet cone will fan out along the beam axis to form a ridge as
it is observed in the RHIC experiments.

In order to be deflected by the soft color fields, the gluon must be radiated
within the medium. Denoting the gluon's momentum along the jet axis
by $p_T$ and its momentum components perpendicular to the jet  by
${\mathbf l}_\perp$, the formation time of the gluon is $t_{\rm f} \sim p_T/{\mathbf l}_{\perp}^2$. 
Therefore, the fields will most
strongly affect those gluons which carry a small fraction of the jet's
longitudinal momentum (with short formation times) and are 
emitted at a substantial angle with respect
to the jet axis. These are precisely the characteristics of the gluons radiated
due to the scattering of the original parton inside the medium.

\begin{figure}[htb!]
\resizebox{2.2in}{1.2in}{\includegraphics[0in,0in][7in,4.6in]{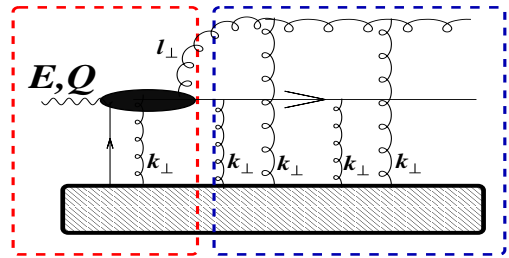}}
    \caption{ A hard scattering followed by multiple soft re-scattering diagram }
    \label{fig1}
\end{figure}

Consider a hard scattering event
with induced radiation as sketched in Fig.~\ref{fig1}.  A jet initiating 
parton is produced when an incoming light-like parton is struck by a virtual 
state with forward energy $E$ and virtuality $Q^2$.  The final state 
parton may be produced off-shell by up to $Q^2$. In cases where it is off-shell, 
it may reduce its virtuality by radiating 
a gluon immediately. It may be produced closer to its on-shell condition and 
re-scatter multiple times prior to radiating 
the gluon. Following the radiation, both remnants of the original parton 
will multiply scatter off  the soft color fields present in the medium. 
The amplitudes for multiple scattering of the hard parton before and after 
the emission of the gluon, as well as the scattering of the radiated gluon 
itself will lead to a destructive interference in the forward collinear 
region of the gluon phase space, the Landau-Pomeranchuck-Migdal (LPM) effect.
This interference suppresses the production of very collinear radiation with 
long formation lengths. As a result the final radiation is produced over a short 
distance and then decoheres from the initiating parton. 
The yield of inclusive gluon radiation in this setting has been studied extensively~\cite{Baier:1996kr,Zakharov:1997uu,Gyulassy:2000er,Wiedemann:2000za,Wang:2001if}. 
Here, we are interested in the more exclusive probability of emission of a gluon 
with large transverse momentum relative to the parent parton. 

To estimate this distribution, we adopt a simplified factorized approach 
illustrated in Fig.~\ref{fig1}. The short distance process which describes the 
production of the parton and its multiple scattering in the dense matter followed 
by emission of the hard gluon is represented by the dark blob. It includes
the two processes commonly referred to as hard-hard and hard-soft double 
scattering~\cite{Wang:2001if} and whose interference leads to the LPM effect for very collinear 
radiation~\cite{Wang:2001if,Guo:2000nz}. 
The subsequent soft scattering of the gluon and the parent parton in 
the medium is treated as independent from the production and radiation process. 
The factorization is indicated by the rectangular boxes in Fig.~\ref{fig1}. 

The remnant jet and the radiated hard gluon then attempt to exit the medium prior 
to fragmenting into a jet  of  hadrons. On their way out, these may 
still endure multiple soft scattering off the dense gluonic field 
of the medium. Such space-like exchanges are most often too soft to 
induce further radiation, however, the  cumulative effect of many 
such scatterings may lead to a non-negligible deflection away from 
the direction of propagation.  While such further re-scattering has a 
diminished role in the modification of  the fragmentation function of the 
parton~\cite{Guo:2006kz},
its additive effect on the relative transverse momentum 
of the radiated gluon with respect to the leading parton is more pronounced. 

To obtain the triggered gluon distribution before the final-state broadening, 
we take the ratio of the cross section $\sigma_{qg}$ for radiating a gluon with 
momentum $ (p_{T_2},{\mathbf l}_\perp)$ in the short distance process depicted in 
the left-hand square in Fig.~\ref{fig1}, and the inclusive cross section 
$\sigma_q$ for production of quark with transverse momentum $p_{T_1}$ and 
rapidity $y$ .  This differential triggered distribution at a fixed impact 
parameter ${\mathbf b}$ and transverse location ${\mathbf r}$ is given as 
Ref.~\cite{Majumder:2005sw},
\begin{eqnarray}
 \frac{d^2 \sigma_{qg} (p_{T_2},{\mathbf l}_\perp )}{d \sigma_q}
&=& 
C(p_{T_1},p_{T_2}) \frac{\alpha_s}{2\pi} \frac{1}{l_\perp^4}  \label{gluon_dist}\\
&\times& \!\!\!\!\!\int_{0}^{\zeta_{max}({\mathbf r})}\!\!\!\!\!\!\!\!\!\!\!\!\!\!d\zeta
\rho({\mathbf r} + \hat{n} \zeta) \frac{\zeta_0}{\zeta} [2 - 2 \cos(\eta_L \zeta)]. \nonumber
\end{eqnarray}
\noindent
In the above equation, $\zeta$ is the distance traveled by the produced 
partons from the primary vertex along the jet axis $\hat{n}$ prior to  
the scattering in the medium. The maximum allowed value $\zeta_{max}$  
for $\zeta$ is the distance from ${\mathbf r}$ to the surface. The factor 
$\eta_L = l_\perp^2/(2p_{T_2}z_1)$ is the inverse of the formation 
time of a gluon with momentum $l_{\perp}$ transverse to the jet axis. 
The final quark momentum fraction is $z_1 = p_{T_1}/(p_{T_1} + p_{T_2})$.
The overall constant $C(p_{T_1},p_{T_2})$  accounts for the fact that the 
final parton momentum in the numerator and denominator of (\ref{gluon_dist}) 
is different. 

The gluon distribution of Eq.~\eqref{gluon_dist} is now used as the 
starting point $\bar{f}(t=t_0)$ for the time evolution of the cone
of radiated gluons around the leading parton due to the influence of 
turbulent color fields. Note, that this distribution includes solely 
those gluons whose production is induced by scattering in the medium 
and does not include any vacuum contribution to the radiation cone.
In order to calculate the deflection of the gluons by turbulent color 
fields, we employ a Fokker-Plank equation:
\begin{equation}
\left[ \frac{\partial}{\partial t} + {\mathbf v}\cdot\nabla_r 
- \nabla_p D({\mathbf p},t)\nabla_p \right] \bar{f} = C[\bar{f}],  
\label{eq:FP-eq}
\end{equation}
with the average parton phase space distribution $f({\mathbf p},{\mathbf r},t)$ and 
the diffusion tensor~\cite{Asakawa:2006tc,Asakawa:2006jn}
\begin{equation}
D_{ij} = \int_{-\infty}^t dt' \left\langle 
  F_i(\bar{r}(t'),t') F_j(\bar{r}(t),t) \right\rangle .
\label{eq:D-def}
\end{equation}
Here ${\mathbf F}=g Q^a ({\mathbf E}^a + {\mathbf v}\times{\mathbf B}^a)$ 
is the color Lorentz force generated by the turbulent color fields, and 
$C[f]$ denotes the collision term. $D_{ij}$ is related to the usual 
transport coefficient for radiative energy loss (jet quenching coefficient)
$\hat{q}$ via:
\begin{equation}
\hat{q} = -2g^{ij}D_{ij}.
\end{equation} 
\noindent
The above equation may be interpreted as a generalization of the 
scalar transport coefficient $\hat{q}$ to a tensor $\hat{q}_{\alpha \beta}$. 
The tensorial form does not influence the total energy lost. However, it 
does influence the angular distribution of emitted radiation. Such an effect is 
not included in Eq.~\eqref{gluon_dist} and will be treated in a forthcoming
publication. The current effort  explores the effect of the anisotropy of 
$D_{ij}$ on the radiated gluon after its decoherence from the leading parton.

For random  transversely polarized color-magnetic fields, the 
diffusion term may be expressed in a simplified form~\cite{Asakawa:2006jn}:
\begin{equation}
\mbox{}\!\!\!\!\! \nabla_p D({\mathbf p}) \nabla_p 
= \frac{ - g^2 Q^2        \langle B^2 \rangle  \tau_{\rm m}          }{2(N_c^2-1)E_p^2} 
    \left[L^{(p)}_{\perp}\right]^2 = \gamma(p)   \left[L^{(p)}_{\perp}\right]^2 , 
\label{eq:D-1}
\end{equation}
where $N_c=3$ is the number of colors, the index $\perp$ denotes the 
vector components transverse to the beam axis, and $\tau_{\rm m}$ is
the autocorrelation length (or time) of the color-magnetic field along
the trajectory of the parton. $E_p$ is the energy of a parton with 
momentum ${\mathbf p}$ and ${\mathbf v} = {\mathbf p}/E_p$ its velocity. 
${\mathbf L}^{(p)} = -i{\mathbf p}\times\nabla_p$ denotes the generator of 
rotations in momentum space.

In the case of interest here, the distribution $f({\mathbf p},t_0)$ 
represents the gluons in the radiation cone of the primary hard parton, 
which have already decohered from the parent parton, as described above. 
Given the form of the diffusion term, the evolution equation is best 
solved by decomposing the momentum distribution function in terms of 
spherical harmonics:
\begin{equation}
f({\mathbf p},t) = \sum_{l,m}  a_{lm}(p,t) Y_m^l (\theta_{{\mathbf p}}, 
\phi_{{\mathbf p}}).
\label{eq:f}
\end{equation}
In the above equation, $l,m$ denote the total and $z$-component of the 
angular momentum in momentum space.  The evolution equations of the 
various components $a_{lm}$, obtained by substituting (\ref{eq:f}) into 
(\ref{eq:FP-eq}), is given by 
\begin{eqnarray}
a_{lm}(p,t) = a_{lm}(p,t_0) e^{-\gamma(p) [l(l+1) - m^2](t - t_0)}.
\end{eqnarray}
Only the spherically symmetric mode $l=m=0$ is unaffected by the evolution, 
while all other modes decay in time. Among these, modes with $m=0$ for a 
given $l$ decay faster than those with $m\neq 0$. As a result, the 
distribution begins to broaden in the $z$-direction (beam direction). 
The evolution time is limited by the depth at which the partons were 
produced and is thus finite. Partons produced at greater depth therefore 
experience greater broadening along the beam direction than those produced 
closer to the surface.  

\begin{figure}[htb!]
{\includegraphics[width=0.85\linewidth]{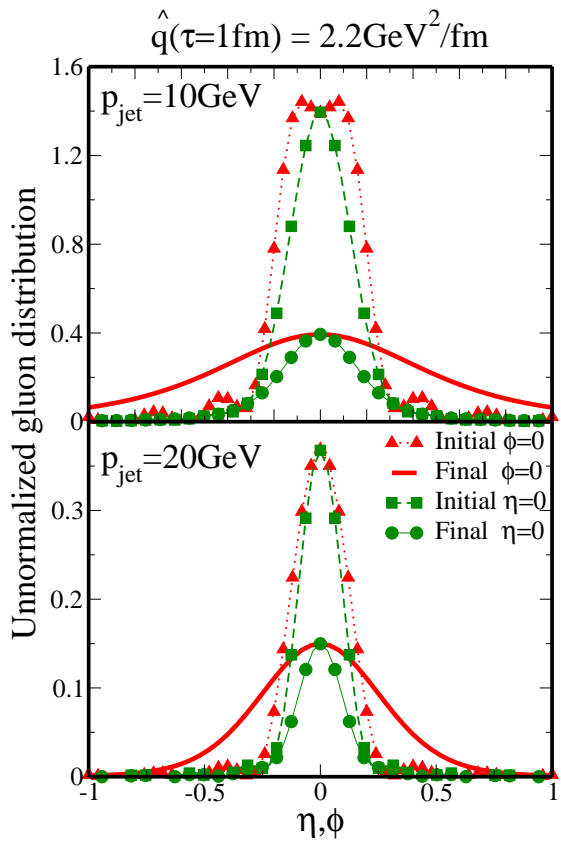}}
    \caption{The gluon distribution around the trigger quark in $\eta$ at $\phi=0$ 
and in $\phi$ at $\eta=0$. The dotted and dashed lines show the 
initial distributions obtained after the gluons have been radiated due to 
interactions of the hard scattered parton with the medium. The solid lines show
the distributions at the moment when the gluons escape from the medium. The 
curves in the top frame correspond to an original jet momentum of 10 GeV/c and 
a final gluon momentum fraction of $z_2 = 0.4$. Those in the lower frame correspond 
to a jet momentum of 20 GeV/c and the same $z_2$.}
    \label{fig2}
\end{figure}

The results of our calculations are presented in Fig.~\ref{fig2}. These 
show the unnormalized associated gluon distribution around 
a jet initiating quark as a function of rapidity $\eta$ for a fixed azimuthal 
angle $\phi=0$, and as a function of $\phi$ at fixed $\eta =0$. The dashed
(dotted) lines show the input distribution ${\bar f}(t_0)$ obtained from 
expression (\ref{gluon_dist}); the solid lines show the modified distributions
at the moment when the gluons leave the medium. Both plots correspond to the 
original jet vertex formed at a depth of 3 fm. The plots in the top part of 
Fig.~\ref{fig2} correspond to an original jet transverse momentum 
$p_{T_1}+p_{T_2} = 10$ GeV/c, while those in the lower part correspond to 
an original jet momentum of $p_{T_1}+p_{T_2} = 20$ GeV. In both cases, the 
gluon carries a forward momentum fraction of $z_2 = 0.4$. 
We assumed in our calculation that the soft gluon density in the medium 
varies with time in accordance with the boost invariant longitudinal expansion 
of an ideal ultrarelativistic liquid. The initial value of the transport 
coefficient associated with the intensity of turbulent color fields was taken 
to be $\hat{q} = 2.2$  GeV$^2/$fm. As is clearly visible in Fig.~\ref{fig2}, 
the associated gluon distribution is considerably broadened in $\eta$ but not
in $\phi$.  The extent of the broadening depends on the energy of the parton 
and drops with increasing energy. 

In summary, we have shown that near side energy loss in terms of radiated gluons 
in the presence of turbulent color fields can explain 
the experimentally observed longitudinal broadening of jet cones in heavy-ion
collisions. 
Our model predicts that this broadening decreases with increasing energy of 
the associated parton (the radiated gluon which fragments to produce the associated hadron). 
An extraction of the transport coefficient $\hat{q}$ from a measurement of such 
broadening in experiment will depend on the energy of the radiated gluon and, 
by extension, on the hadronization mechanism. The momentum range 
of the partons contained in the rapidity broadened jet cone is in the regime 
where hadronization occurs dominantly via parton recombination. We therefore
expect an enhanced baryon to meson ratio of hadrons contained in the ridge.

As mentioned earlier, turbulent color fields generated by Weibel instabilities 
in the early quark-gluon plasma have recently been proposed as a mechanism 
for the early thermalization of the matter produced at RHIC and as the source 
of a small anomalous viscosity.  The current work has identified the broadening 
of triggered jet cones in rapidity as a more directly observable manifestation 
of the presence of turbulent color fields in the matter.  If our interpretation can be
confirmed by additional measurements, it will demonstrate the existence of 
plasma instabilities generating these turbulent fields and establish their role 
in the transport of hard probes through a quark-gluon plasma.

{\it Acknowledgments:} This work was supported in part by a grant 
from the U.~S.~Department of Energy (DE-FG02-05ER41367).

\end{document}